\documentclass[1pt,a4paper]{article}
\usepackage{type1cm,amsmath,indentfirst}
\usepackage{amssymb}
\numberwithin{equation}{section}
\usepackage{graphicx}
\begin{document}
%%% Title page %%%%%
\begin{titlepage}

 \renewcommand{\thefootnote}{\fnsymbol{footnote}}
\begin{flushright}
 \begin{tabular}{l}
 %DESY 10-098\\
 %WITS-CTP-54%\\
% arXiv:1209.xxxx\\ %This should be replaced after submittion.
% \today %This should be commented out.
 \end{tabular}
\end{flushright}

 \vfill
 \begin{center}

% \vskip 2.5 truecm

\noindent{\large \textbf{$\mathcal{N}=1$ supersymmetric higher spin holography on AdS$_3$
%(temp.)
}}\\
\vspace{1.5cm}

\noindent{ Thomas Creutzig,$^{a,b}$\footnote{E-mail: tcreutzig@mathematik.tu-darmstadt.de} Yasuaki Hikida$^c$\footnote{E-mail:
hikida@phys-h.keio.ac.jp} and Peter B. R\o nne$^{d}$\footnote{E-mail: peter.roenne@uni-koeln.de}}
\bigskip

 \vskip .6 truecm
\centerline{\it $^a$Fachbereich Mathematik,
Technische Universit\"{a}t Darmstadt,}
\centerline{\it Schlo\ss gartenstr. 7,
64289 Darmstadt, Germany}
\medskip
\centerline{\it $^b$Hausdorff Research Institute for Mathematics,}
\centerline{\it{Poppelsdorfer Allee 45,
53115 Bonn,
Germany }}
\medskip
\centerline{\it $^c$Department of Physics, and Research and Education
Center for Natural Sciences,}
\centerline{\it  Keio University, Hiyoshi, Yokohama 223-8521, Japan}
\medskip
\centerline{\it $^d$Institut f\"{u}r Theoretische Physik, Universit\"{a}t zu K\"{o}ln,
} \centerline{\it
Z\"{u}lpicher Stra{\ss}e 77, 50937 Cologne, Germany}
 \vskip .4 truecm

 \end{center}

 \vfill
\vskip 0.5 truecm

\begin{abstract}

We propose a duality between a higher spin ${\cal N}=1$ supergravity on AdS$_3$ and
a large $N$ limit of a family of ${\cal N}=(1,1)$ superconformal field theories.
The gravity theory is  an ${\cal N}=1$ truncation of the ${\cal N}=2$ supergravity found by Prokushkin
and Vasiliev, and the dual conformal field theory is defined by a supersymmetric coset model.
We check this conjecture by comparing one loop partition functions and find
agreement. Moreover, we study the symmetry of the dual coset model and in particular compute fields of the coset algebra of dimension $3/2,2,2$ and $5/2$ explicitely.

\end{abstract}
\vfill
\vskip 0.5 truecm

\setcounter{footnote}{0}
\renewcommand{\thefootnote}{\arabic{footnote}}
\end{titlepage}

\newpage

\tableofcontents
%%%%%%%%%%%%%%%%%%%%%%%%%%%%%%%%%%%%%%%%%%%%%%%%%%%%%%%%%%%%%%%%%%%%%%

\section{Introduction}

Recently, holography involving higher spin gauge theories has received a lot of attention.
Higher spin gauge theories are believed to be related to the tensionless limit of superstring theory,
and indeed higher spin holography may be seen as a simplified, but non-trivial, version of AdS/CFT duality in superstring theory.
A famous example is the proposal by Klebanov and Polyakov in \cite{Klebanov:2002ja}
which says that a 4 dimensional higher spin gauge theory
is dual to the large $N$ limit of the O$(N)$ vector model. Two years ago it was conjectured in \cite{GG}
that  a 3 dimensional higher spin gauge theory is dual to a large $N$ minimal model,
see \cite{Gaberdiel:2012uj} for a review.
There are several generalizations of this duality; a truncated version was considered in
\cite{Ahn:2011pv,GV}, and the full ${\cal N}=2$ supersymmetric version was introduced
in \cite{CHR}.
In this paper we would like to propose and test the ${\cal N}=1$ supersymmetric
version of the duality.

The higher spin gauge theories appearing in these dualities are truncated versions of the
${\cal N}=2$ supergravity by Prokushkin and Vasiliev \cite{PV1}.
A bosonic truncation is used in the original proposal \cite{GG} which has
an infinite series of higher spin gauge fields with spins $s=2,3,\ldots$ and two complex scalars.
The dual theory is a large $N$ limit of the minimal model with higher spin W$_N$
symmetry \cite{Bouwknegt:1992wg}.
Currently there is much evidence for the duality based on the symmetry
\cite{Henneaux:2010xg,Campoleoni:2010zq,Gaberdiel:2011wb,Campoleoni:2011hg,Gaberdiel:2012ku}
and the spectrum \cite{Gaberdiel:2011zw}, and also correlators have been calculated \cite{CY,Papadodimas:2011pf,Ahn:2011by,AKP,Chang:2011vk}.
In the case of finite $N$, a refined version of the duality has been proposed in \cite{Gaberdiel:2012ku}, however, we will here only be concerned with the strict large $N$ limit.
In \cite{Ahn:2011pv,GV} a further truncation is used for the gravity theory,
and it then includes even spin gauge fields with $s=2,4,6,\ldots$ and two real scalars
with the mass  $M^2 = - 1 + \lambda^2$.
The dual theory is the large $N$ limit of the WD$_N$ minimal model described by the coset theory%
\footnote{
As pointed out in \cite{Ahn:2011pv,GV},
the WB$_N$ minimal model is a candidate as well.}
\begin{align}
\label{bosoncoset}
 \frac{\widehat{\text{so}}(2N)_k \oplus \widehat{\text{so}}(2N)_1}{\widehat{\text{so}}(2N)_{k+1}} ~.
\end{align}
In the limit we also take $k$ to infinity, but fix the 't Hooft parameter
\begin{align}
\label{bosonpara}
 \lambda = \frac{2N}{2N+k-2} ~,
\end{align}
which is identified with the $\lambda$ parameter in the mass of the bulk scalars.
This conjecture was supported by the analysis of RG-flow in \cite{Ahn:2011pv}
and the comparison of one-loop partition functions in \cite{GV}, see also \cite{Ahn:2012gw} for work on the spin-4 operator in correlators.

In \cite{CHR} we extended the duality to the case with supersymmetry.
The untruncated version of the ${\cal N}=2$ supergravity  in \cite{PV1} is proposed to be
dual to the ${\cal N}=(2,2)$ $\mathbb{C}\text{P}^N$ Kazama-Suzuki model
\cite{Kazama:1988qp,Kazama:1988uz}.
The partition functions of the gravity theory and the CFT are shown to match in
\cite{CG}, and the symmetry algebras are analyzed in
\cite{Henneaux:2012ny,Hanaki:2012yf,Ahn:2012fz,Candu:2012tr}.
Other related works may be found in \cite{Fredenhagen:2012rb,Ahn:2012vs,Tan:2012xi,Datta:2012km,Fredenhagen:2012bw,CHR2}.
By including supersymmetry, quantum effects are known to become more tractable in general.
Moreover, the relation to superstring theory could be more transparent as was also mentioned
in \cite{Henneaux:2012ny}.
In \cite{PV1} several ${\cal N}=1$ truncations have been also discussed, and we would like
to consider the simplest one without any matrix degrees of freedom.
The supergravity includes both bosonic and fermionic higher spin gauge fields and
also massive scalars and fermions, as in the untruncated case.
Our proposal is that the dual CFT is given by the ${\cal N}=(1,1)$ super coset
\begin{align}
\label{susycoset}
 \frac{\widehat{\text{so}}(2N+1)_k \oplus \widehat{\text{so}}(2N)_1}{\widehat{\text{so}}(2N)_{k+1}} ~.
\end{align}
There is no enhancement of supersymmetry from ${\cal N}=1$ to
${\cal N}=2$ since this coset does not satisfy the condition in
\cite{Kazama:1988qp,Kazama:1988uz}\footnote{The condition for enhancement to $\mathcal N=2$ superconformal symmetry
is that the coset manifold $G/H$ is a Hermitian symmetric space. This is not true for $\text{SO}(2N+1)/\text{SO}(2N)$ with $N>1$.}. We need to take the large $N$ limit
with 't Hooft parameter%
\footnote{For large $N,k$, the difference between \eqref{bosonpara} and \eqref{susypara} is irrelevant.}
\begin{align}
 \lambda = \frac{2N}{2N + k -1}
 \label{susypara}
\end{align}
kept finite. Again, this parameter is identified with the one in the masses of the bulk matter.
We will underpin this conjecture by showing that
the supergravity  and the CFT partition functions match, i.e. the spectrum is the same on both sides. Further, we will also study the symmetry of the super coset model.

This paper is organized as follows:
In the next section we give explicit formulas for one-loop partition functions in the
higher spin gravity theories.
In section \ref{bosonic}, the one-loop partition function of truncated
bosonic gravity theory is reproduced by the 't Hooft limit of the coset model
\eqref{bosoncoset}. This was already done in \cite{GV}, but we obtain the
same result by using the different method adopted in \cite{CG} as a
preparation of the later analysis.
In section \ref{susy}, we introduce the  ${\cal N}=(1,1)$ super coset \eqref{susycoset}
and study its torus partition function. We show that in the 't Hooft limit it
reproduces the one-loop partition function of the ${\cal N}=1$ supergravity.
In section \ref{symmetry}, we study the symmetry of the super coset model
\eqref{susycoset}.  Section \ref{conclusion} is devoted to conclusion and discussions.
In appendix \ref{N=1}, we review the ${\cal N}=1$ truncation of the ${\cal N}=2$
supergravity found in \cite{PV1} and show that the ${\cal N}=1$ higher spin algebra
is the analytical continuation of the osp$(2N+1|2N)$ Lie superalgebra.
In appendix \ref{so}, we summarize useful properties of orthogonal Lie algebras.

\section{Higher spin gravity theories}
\label{gravitypf}

In this section, we study truncated versions of the higher spin $\mathcal{N}=2$ supergravity in \cite{PV1}
and obtain explicit formulas for their one-loop partition functions.
In the first subsection, we review the bosonic gravity theory having only gauge fields of even spin
$s=2,4,6,\ldots$ which is proposed in \cite{Ahn:2011pv,GV} as the gravity dual of
the WD$_N$ minimal model.
In subsection \ref{N=1g}, we consider the larger ${\cal N}=1$ truncation of ${\cal N}=2$
supergravity via an anti-automorphism which was also introduced in \cite{PV1}, and which we review in appendix \ref{N=1}.

\subsection{The bosonic truncation}
\label{bosong}

In \cite{GG} a bosonic truncation of the ${\cal N}=2$ supergravity introduced in \cite{PV1} is utilized to
construct a simplified version of the AdS/CFT correspondence.
The gravity theory includes massless gauge fields with spins $s=2,3,4,\ldots$ and two
complex scalars with mass
\begin{align}
M^2 = - 1 + \lambda^2 ~.
\label{massGG}
\end{align}
The massless sector can be described by a Chern-Simons gauge theory based on the
hs$[\lambda]$ Lie algebra which can be reduced to sl$(N)$
for $\lambda=\pm N$. The dual CFT is proposed to be a large $N$ limit of the
WA$_N$ minimal model.

The WD$_N$ (and WB$_N$) version of \cite{GG} was treated by \cite{Ahn:2011pv} and \cite{GV} (the last reference mostly on the former algebra).
The bulk side is a truncated version of the hs$[\lambda]$ algebra containing only even spins,
see \cite{GV,CGKV}.
The one-loop
partition function for the gauge sector is given by \cite{Giombi:2008vd,David:2009xg,Gaberdiel:2010ar}
\begin{align}
 Z_\text{gauge} = \prod_{l=1}^\infty Z^{(2l)}_\text{gauge}
 =  \prod_{l=1}^\infty\prod_{n=2l}^{\infty} \frac{1}{|1 - q^n|^2} ~,
\end{align}
where $q=\exp(\tau)$ is a modulus for the boundary torus of the Euclidean AdS$_3$.
The bulk side contains two real scalars with the same mass as in eq. \eqref{massGG}.
The fall-off behaviour at the boundary is chosen oppositely for the two scalars like in the WA$_N$ case. In the dual boundary CFT this gives two real scalars with conformal weights
\begin{align}
 h_+=\frac{1+\lambda}{2}\ ,\qquad  h_-=\frac{1-\lambda}{2}\ .
 \label{dcw}
\end{align}
For each scalar field with the dual conformal
weight $h$, the partition function is \cite{Giombi:2008vd,David:2009xg}
\begin{align}
 Z_\text{scalar}^h = \prod_{m,n=0}^\infty \frac{1}{1 - q^{h+m} \bar q^{h+n}} ~.
\end{align}
The one-loop partition function of the gravity theory is then given by
\begin{align}
Z^\lambda_\text{1-loop} = Z_\text{gauge} Z_\text{scalar}^{h_+}
 Z_\text{scalar}^{h_-} ~.
\end{align}

\begin{figure}
 \begin{center}
\includegraphics[scale=0.7]{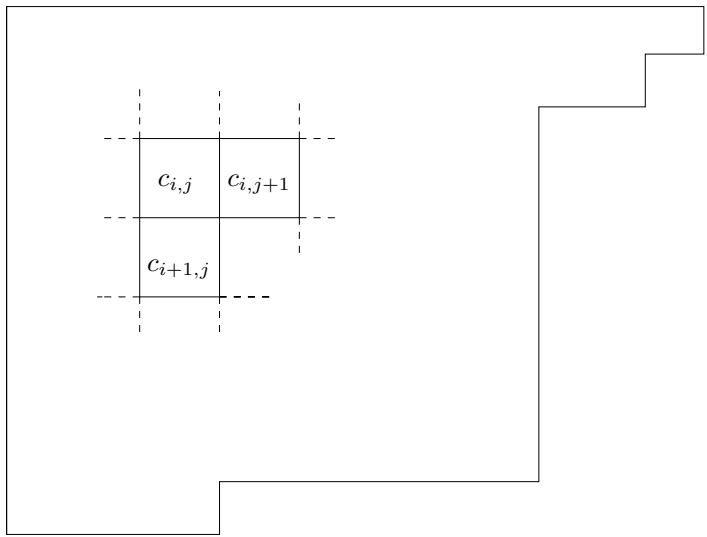}
\put(-205,132){${c_{i,j}}$}
\put(-209,100){$c_{i+1,j}$}
\put(-179,132){$c_{i,j+1}$}
 \end{center}
 \caption{
A Young tableau Tab$_\Lambda$ of a shape $\Lambda$. In each box of the Young diagram $\Lambda$,
 we assign a non-negative number $c_{i,j}$ with a rule that $c_{i,j} \leq c_{i,j+1}$ and $c_{i,j} < c_{i+1,j}$. A Young supertableau STab$_\Lambda$ of a shape $\Lambda$ are also given by a Young diagram $\Lambda$
and a non-negative number $c_{i,j}$ in a each box. However, the rules for $c_{i,j}$ are a bit different. The numbers should always satisfy the conditions $c_{i,j} \leq c_{i,j+1}$ and $c_{i,j} \leq c_{i+1,j}$. Further
$c_{i,j} < c_{i,j+1}$ if $c_{i,j}$ and $c_{i,j+1}$ are odd, and  $c_{i,j} < c_{i+1,j}$ if $c_{i,j}$ and $c_{i+1,j}$
are even.}
 \label{fig:one}
\end{figure}

In order to compare the gravity partition function to the dual CFT partition function,
it is convenient to rewrite the partition function of the matter sector.
We introduce a Young tableau Tab$_\Lambda$ of shape $\Lambda$.
Here we assign a non-negative integer $c_{i,j}$ to the box in the Young diagram $\Lambda$
at the $i$-th row and the $j$-th column, see figure \ref{fig:one}.
The rules for the numbers  $c_{i,j}$ are that the entries do not decrease along a row and increase along a column.
Then the partition function of a scalar field can be rewritten as \cite{Gaberdiel:2011zw,CG}
\begin{align}
 Z_\text{scalar}^h =   \sum_{\Lambda}
  | \text{ch}_\Lambda (U(h)) |^2 ~,
\end{align}
where the character of the representation $\Lambda$ is defined as
\begin{align}
 \text{ch}_\Lambda (U(h)) = \sum_{T \in \text{Tab}_\Lambda} \prod_{j \in T} q^{h+j} ~,
 \qquad U(h)_{jj}=q^{h+j} ~.
 \label{chL}
\end{align}
The one-loop partition function of the gravity theory is thus summarized as
\begin{align}
 Z^\lambda_\text{1-loop} = Z_\text{gauge} \sum_{\Lambda , \Xi}
  | \text{ch}_\Lambda (U(h_+)) \text{ch}_\Xi (U(h_-)) | ^2 \label{gpf}
\end{align}
with $h_\pm$ given in \eqref{dcw}.

\subsection{The $\mathcal{N}=1$ truncation}
\label{N=1g}

In \cite{PV1} the truncation from $\mathcal{N}=2$ to $\mathcal{N}=1$ supergravity has been discussed,
see also appendix \ref{N=1}. The ${\cal N}=1$ theory also has massless higher spin gauge fields and
matter fields. Let us start from the massless sector.
As derived in appendix  \ref{N=1}, the massless sector can be described by a large $N$ limit of
 $\text{osp}(2N+1|2N)  \oplus \text{osp}(2N+1|2N) $ Chern-Simons gauge  theory.%
 \footnote{It might be possible to use $\text{osp}(2N-1|2N)  \oplus \text{osp}(2N-1|2N) $ Chern-Simons gauge  theory.}
In order to see the spin content of the theory, we have to identify an osp$(1|2)$ subalgebra
as the ${\cal N}=1$ supergravity sector.
As in \cite{CHR}, we adopt
the superprincipal embedding of osp$(1|2)$, which  gives us \cite{Frappat:1992bs}
\begin{align}
 \textrm{osp}(2N+1|2N)=\sum_{k=1}^{N} \left(R_{2k-1}\oplus R_{2k-1/2}\right)\ .
\end{align}
Here $R_j$ is the representation of osp$(1|2)$ decomposing under the sl(2) as
$R_j=D_j\oplus D_{j-1/2}$, where $D_j$ denotes the $2j+1$ dimensional representation of sl(2).
This gives a spin decomposition of the form
\begin{align}
 \textrm{osp}(2N+1|2N)=\sum_{k=1}^{N} \left(D_{2k-1/2}\oplus 2 D_{2k-1}\oplus D_{2k-3/2}\right)\ .
\end{align}
We see that we have no odd integer spins.
In the infinite $N$ limit we have two fields of each even spin, and one field of each half odd integer spin.

The spin $j$ of the embedded sl(2) is related to the space-time spin as $s=j+1$, and a more close examination
shows that the integer and the half-integer spin components are bosonic and fermionic elements of
osp$(2N+1|2N)$, respectively.
With the help of the results for ${\cal N}=2$ supergravity in \cite{CHR}, we can write down
the contribution from higher spin fields as
\begin{align}
 {\cal Z}_\text{gauge}   =  \prod_{l=1}^\infty (Z^{(2l)}_B)^2 \prod_{s=2}^\infty Z^{(s-1)}_F~,
\end{align}
where
\begin{align}
Z^{(s)}_B = \prod_{n=s}^{\infty} \frac{1}{|1 - q^n|^2} \ ,\qquad
Z^{(s)}_F = \prod_{n=s}^{\infty} |1 + q^{n+\frac12}|^2 \
\end{align}
are the partition functions of a bosonic field with spin $s$ and a fermionic field with spin $s+1/2$,
respectively.

One advantage of the Chern-Simons description is that we can easily read off the classical asymptotic symmetry
near the boundary of AdS$_3$ \cite{Henneaux:2010xg,Campoleoni:2010zq,Gaberdiel:2011wb,Campoleoni:2011hg}.
The Chern-Simons theory is a topological theory, and dynamical degrees of freedom exist only at the boundary.
The boundary degrees can be described by $\text{osp}(2N+1|2N)$ Wess-Zumino-Novikov-Witten
model, whose symmetry is the affine $\text{osp}(2N+1|2N)$ Lie superalgebra. For the application to the AdS/CFT
correspondence, we have to assign the boundary conditions ensuring the bulk space being asymptotically AdS space.
It was shown in \cite{Campoleoni:2010zq,Campoleoni:2011hg} that this condition is the same
as for the classical Hamiltonian reduction (see, for instance, \cite{Bouwknegt:1992wg}). Thus the classical
asymptotic symmetry of the ${\cal N}=1$ truncated theory is obtained by the
Hamiltonian reduction of  affine $\text{osp}(2N+1|2N)$ Lie superalgebra in a large $N$ limit.

The matter sector consists of a single $\mathcal{N}=1$ hypermultiplet, having two complex scalars with masses respectively
\begin{align}\label{massCHRB}
 ( M^B_-)^2 = - 1 + \lambda  ^2 \ , \qquad( M^B_+ )^2 = - 1 + (\lambda - 1 )^2 ~,
\end{align}
 and two fermions with mass
\begin{align}\label{massCHRF}
 (M^F_\pm)^2 = (\lambda - \tfrac{1}{2})^2 ~.
\end{align}
Also for the fermions we can choose two types of boundary conditions.
We choose these such that
the boundary conformal dimensions simply become a real version of the $\mathcal{N}=2$ case \cite{CHR} (see also \cite{CG})
\begin{align}\label{eq:bulkconformaldimension}
 (\Delta^B_+  , \Delta^F_{\pm} , \Delta^B_-) = (2 - \lambda , \tfrac{3}{2} - \lambda , 1 - \lambda )  , ~
 (\lambda , \tfrac{1}{2} + \lambda , 1 + \lambda)\ .
\end{align}
The one-loop partition function of  the matter part is \cite{CHR}
\begin{align}
{\cal Z}_\text{matter} = {\cal Z}_\text{hyper}^{\frac{\lambda}{2}}
{\cal Z}_\text{hyper}^{\frac{1-\lambda}{2}}
~, \qquad
{\cal Z}_\text{hyper}^h = Z^{h+\frac12}_\text{scalar} ( Z_\text{spinor}^{h } )^2 Z_\text{scalar}^{h}
\end{align}
where (note no square in the bosonic partition function)
\begin{align}
Z_\text{scalar}^{h} = \prod_{l,l'=0}^\infty \frac{1}{1 - q^{h+l} \bar q^{h+l'}} \ , \qquad
(Z_\text{spinor}^{h })^2 = \prod_{l,l'=0}^\infty ( 1 + q^{h+l} \bar q^{h + \frac12 +l'} )
 ( 1 + q^{h + \frac12 +l} \bar q^{h +l'} ) ~. \nonumber
\end{align}
As in the bosonic case, it is convenient to define the supercharacter%
\footnote{This is a character of gl$(\infty|\infty)_+$ considered in \cite{CG}.
The bosonic one \eqref{chL} is the character of gl$(\infty)_+$ which is the Lie algebra of
infinite dimensional matrices. The generators may be given by $ (e_{ij} )_{k,l} =
\delta_{i,l} \delta_{j,k}$ and only those with finite $i,j$ are considered.
The supergroup gl$(\infty|\infty)_+$ is quite similar to gl$(\infty)_+$, but now
$e_{ij}$ is bosonic for even $i+j$ and fermionic for odd $i+j$. }
\begin{align}
 \text{sch}_\Lambda ({\cal U} (h)) = \sum_{T \in \text{STab}_\Lambda}
 \prod_{j \in T}q^{h+\frac{j}{2}}  ~, \qquad
 {\cal U} (h)_{jj} = (-1)^j q^{h+\frac{j}{2}}   ~.
 \label{schL}
\end{align}
Here STab$_{\Lambda}$ represents a Young supertableau of shape $\Lambda$.
In the supertableau, a non-negative integer is assigned to each box of the
Young diagram $\Lambda$ with rules specified in figure \ref{fig:one}.
Then following \cite{CG}, the partition function  of the matter sector can be given as
\begin{align}
{\cal Z}_\text{hyper} ^h =  \sum_\Lambda  |\text{sch}_\Lambda ({\cal U} (h) )|^2\ .
\end{align}
Therefore, the one-loop partition function of the ${\cal N}=1$ truncated theory
can be expressed as
\begin{align}
 {\cal Z} = {\cal Z}_\text{gauge} \sum_{\Lambda ,\Xi}
  |\text{sch}_\Lambda ({\cal U} (h_+)) \, \text{sch}_\Xi ({\cal U} (h_-)) |^2 ~,
  \label{sgpf}
\end{align}
where $h_+ = \frac{\lambda}{2}$ and $h_- = \frac{1-\lambda}{2}$.

\section{Holography for SO($2N$)}
\label{bosonic}

It was proposed in \cite{GG} that a bosonic truncation of Prokushkin-Vasiliev theory \cite{PV1}
is dual to a large $N$ limit of W$_N$ minimal model. A further consistent truncation is
possible in the gravity theory as discussed in section \ref{bosong}, and the dual CFT
is conjectured to be the  WD$_N$ minimal model with the coset description \eqref{bosoncoset}
 \cite{Ahn:2011pv,GV}.
In order to compare with the classical gravity theory, we need to take a large $N$
limit while keeping the 't Hooft coupling \eqref{bosonpara} finite, and
this parameter is identified with $\lambda$ in \eqref{massGG}.
The equivalence of the spectrum in this limit was shown in \cite{GV} by directly applying the
method in \cite{Gaberdiel:2011zw}.
In this section, we will obtain the same result by making use of
the different method in \cite{CG}.
In the next section, we will use the same method to show the matching of the spectrum in the ${\cal N}=1$ supersymmetric version of  the duality.

\subsection{The dual CFT}

We would like to reproduce the gravity partition function \eqref{gpf} from the
viewpoint of the dual coset CFT \eqref{bosoncoset}%
\footnote{We heard that in \cite{CGKV} the duality is refined so as to be applicable for finite $N,k$.
It is pointed out there that an $\mathbb{Z}_2$ orbifold should be used
instead of \eqref{bosoncoset}, but this difference disappears in the large $N$ limit,
like we also found in \eqref{sorbifold}. Further, this subtlety does not arise in the WB$_N$ case.}
\begin{align}
  \frac{\widehat{\text{so}}(2N)_k \oplus \widehat{\text{so}}(2N)_1 }{\widehat{\text{so}}(2N)_{k+1}} ~,
  \label{coset}
\end{align}
 whose central charge is
\begin{align}
 c = N \left[ 1 - \frac{(2N-1)(2N-2)}{(k + 2N - 2) (k + 2 N - 1)}\right] ~.
\end{align}
We use the diagonal embedding of $\widehat{\text{so}}(2N)_{k+1}$ into $\widehat{\text{so}}(2N)_k \oplus \widehat{\text{so}}(2N)_1$,
and
the states of the coset \eqref{coset} are obtained by the decomposition
\begin{align}
 \Lambda \otimes \omega = \oplus_\Xi (\Lambda,\omega ; \Xi ) \otimes \Xi ~.
 \label{dec}
\end{align}
The states are thus labeled by  $(\Lambda,\omega,\Xi)$, which are the highest weights of the
representations of $\widehat{\text{so}}(2N)_k$, $\widehat{\text{so}}(2N)_1$, $\widehat{\text{so}}(2N)_{k+1}$, respectively.
Some basics of so($2N$) Lie algebra may be found in appendix \ref{so}.
For $\widehat{\text{so}}(2N)_1$, there are only four representations: The identity, vector, spinor and cospinor
representations.
The selection rule is
\begin{align}
 \Lambda + \omega - \Xi \in Q_{2N}
\end{align}
where $Q_{2N}$ is the root lattice of so$(2N)$.
The congruence class of so$(2N)$  is $\mathbb{Z}_4$ for $N = 2 n +1$
or $\mathbb{Z}_2 \times \mathbb{Z}_2$ for $N=2 n$ with $n \in \mathbb{Z}$
(see, e.g., \cite{cft}), and this equation uniquely determines $\omega$
given $\Lambda$ and $\Xi$, and we simply denote the states by $(\Lambda;\Xi)$. Moreover, there are field identifications
$(\Lambda;\Xi) \simeq (A \Lambda; A\Xi)$ with an outer automorphism $A$ of the
affine algebra $\widehat{\text{so}}(2N)_k$ and $\widehat{\text{so}}(2N)_{k+1}$ \cite{Gepner:1989jq}.
The conformal weight of the state $(\Lambda ; \Xi)$ is given by
\begin{align}
 h_{(\Lambda ; \Xi)} = \frac{C_{2N} (\Lambda)}{k+2N-2} + \frac{C_{2N}(\omega)}{2N-1}
  - \frac{C_{2N}(\Xi)}{k+2N-1} +n ~,
\end{align}
where $C_{2N} (\Lambda)$ is the quadratic Casimir of so($2N$)
and the integer $n$ is the grade at which $\Xi$ appears in $\Lambda \otimes \omega$.

We define the characters of $\widehat{\text{so}}(2N)_k$ and the branching function as
\begin{align}
 \text{ch}_\Lambda^{2N,k} (q,e^H) = \text{tr}_\Lambda  q^{L_0} e^H ~,
 \qquad
 b_{\Lambda;\Xi}^{2N,k} (q) = \text{tr}_{(\Lambda;\Xi) } q^{L_0} ~,
 \label{bf}
\end{align}
where $L_0$ is the zero mode of energy momentum tensor and $H$
is the Casimir element of so$(2N)$.  From the decomposition  \eqref{dec}
we find
\begin{align}
 \text{ch}_\Lambda^{2N,k} (q,e^{H}) \,  \text{ch}_\omega^{2N,1} (q,e^{H})
 = \sum_\Xi b^{2N,k}_{\Lambda ; \Xi}  (q) \, \text{ch}_\Xi^{2N,k+1} (q,e^{H}) ~.
 \label{bf2}
\end{align}
In addition to the chiral sector, the CFT has an anti-chiral sector.
We consider the charge-conjugated theory with the Hilbert space
${\cal H}= \sum (\Lambda ; \Xi) \otimes \overline{(\Lambda ; \Xi)}$.
We should take care of the field identification when the sum is taken.
The partition function is then given by the diagonal modular invariant
\begin{align}
 Z^{2N,k} (q) = |q^{-\frac{c}{24}}|^2 \sum_{[\Lambda;\Xi]} | b^{2N,k}_{\Lambda ; \Xi}  (q)|^2 ~.
\end{align}

In order to compare with the classical gravity theory, we take a large $N,k$ limit with
the 't Hooft parameter \eqref{bosonpara}
\begin{align}
\lambda = \frac{2N}{2N + k - 2} \nonumber
\end{align}
kept finite.
Since representations of order $N^2$ decouple in this limit,
we keep those whose conformal weights are of order $N$.
With this criterion, the highest weights for the representations
that survive can be labeled by Young tableaux \cite{GV} (see also appendix \ref{so}).
Moreover, representations are self-conjugate for the orthogonal Lie algebra.
Denoting the number of  boxes in the $i$-th row by $l_i$ and in the $j$-th column by $c_j$,
the quadratic Casimir is expressed as \eqref{casimirso}
\begin{align}
 C_{2N}(\Lambda) = |\Lambda| \left(N-\frac12 \right) + \frac12 \left(\sum_{i=1}^{N-2} l_i^2 - \sum_j c_j^2 \right) ~.
\end{align}
We consider Young diagrams with only finitely many boxes in the large $N$ limit,
and the above equation implies that $C_{2N}(\Lambda) \sim N|\Lambda|$
where $|\Lambda|$ is the number of boxes of  the corresponding Young diagram.

\subsection{Comparison of partition functions}

We would like to compute the branching function $b^{2N,k}_{\Lambda;\Xi}(q)$ in eq. \eqref{bf}
in the 't Hooft limit. For $q$ having real part less than one, we can neglect
$q^N$ and $q^k$.
Following \cite{GV} we first study the large $k$ behavior of the characters of
$\widehat{\text{so}}(2N)_k$ and then compute the branching function utilizing \eqref{bf2}.
For large $k$,  the character becomes \cite{Bouwknegt:1992wg,Gaberdiel:2011zw,CG}
\begin{align}
  \text{ch}_\Lambda^{2N,k} (q,e^{H}) \simeq
  \frac{q^{h_\Lambda^{2N,k}} \text{ch}_\Lambda^{2N} (e^H)}{\prod_{n=1}^\infty [ (1 - q^n)^N
    \prod_{\alpha \in \Delta_{2N}} ( 1 - q^n e^{\alpha (H) } ) ] } ~,
    \label{KW}
\end{align}
where we have used the Weyl-Kac formula  and $\Delta_{2N}$ denotes the roots of so$(2N)$.
For large $k$, the affine Weyl group reduces to the finite Weyl group as discussed in \cite{Gaberdiel:2011zw},
which leads to the character  of the finite so$(2N)$ Lie algebra ch$_\Lambda^{2N}(e^H)$.
The conformal dimension is now
\begin{align}
 h_\Lambda^{2N,k} = \frac{C_{2N} (\Lambda)}{k+2N-2}  ~.%\simeq \frac{\lambda}{2} |\Lambda|
\end{align}

For large $k$ we can reduce the relation \eqref{bf2} to
\begin{align}
 \text{ch}_\Lambda^{2N} (e^H) \, \text{ch}_\omega^{2N,1} (q, e^H)
 = \sum_{\Xi} a^{2N}_{\Lambda;\Xi} (q) \, \text{ch}_\Xi^{2N} (e^H) ~,
 \label{defa0}
\end{align}
with the help of \eqref{KW}. Here
the $k$-independent function $a^{2N}_{\Lambda;\Xi} (q) $  is related to
the branching function as
\begin{align}
 b^{2N,k}_{\Lambda ; \Xi}  (q) \simeq q^{ h_\Lambda^{2N,k}  - h_\Xi^{2N,k+1} } a^{2N}_{\Lambda;\Xi} (q) ~.
\end{align}
As will be shown now, the  function $a^N_{\Lambda;\Xi} (q) $ can be written as%
\begin{align}
 a_{\Lambda ; \Xi}^{2N} (q) = \sum_{\Pi} N_{\Lambda \Pi}^{(2N)\Xi} a_{0;\Pi}^{2N} (q)
 = \sum_{\Pi} N_{\Lambda \Xi}^{(2N)\Pi}\,  a_{0;\Pi}^{2N} (q)
\end{align}
with $N_{\Lambda \Xi}^{(2N)\Pi}$ being the Clebsch-Gordan coefficients of so$(2N)$.
First notice that
\begin{align}
 \text{ch}_\omega^{2N,1} (q, e^H)
 = \sum_{\Xi} a^{2N}_{0;\Xi} (q) \, \text{ch}_\Xi^{2N} (e^H)  ~, \label{defa}
\end{align}
which is obtained from \eqref{defa0} with $\Lambda = 0$.
Then, we use
\begin{align}
 \text{ch}_\Lambda^{2N} (e^H)  \, \text{ch}_\Pi^{2N} (e^H) = \sum_\Xi  N_{\Lambda\Pi}^{(2N)\Xi} \text{ch}_\Xi^{2N} (e^H)
 \label{cgc}
\end{align}
and the fact that the representations are now self-conjugate.

Next, we take the large $N$ limit.
As discussed in the previous subsection, the highest weight representations can be labeled by
Young diagrams with finitely many boxes in the limit.
In the next subsection, we will obtain
\begin{align}
 &a_{0;0} (q) = \lim_{N \to \infty} a_{0;0}^{2N} (q)  =  \prod_{l=1}^\infty\prod_{n=2l}^{\infty} \frac{1}{1 - q^n }~, \label{a00}\\
 &a_{0;\Xi} (q) = \lim_{N \to \infty} a_{0;\Xi}^{2N} (q)  = a_{0;0} (q)  \,
  \text{ch}_{\Xi^t} (U(\tfrac12))~, \label{a0Xi}
\end{align}
where the character is defined in \eqref{chL}.
In the transposed expression $\Xi^t$, the rows and columns are exchanged.
Moreover, the conformal weight becomes
\begin{align}
 h_\Lambda^{2N,k} \simeq \frac{\lambda}{2} |\Lambda| ~.
\end{align}
Using these facts, the branching function in the 't Hooft limit can be found as
\begin{align}
 b^\lambda_{\Lambda;\Xi} (q) = q^{\frac{\lambda}{2} (|\Lambda| - |\Xi|)} a_{0;0} (q)
 \sum_{\Pi} N_{\Lambda \Xi}^\Pi \, \text{ch}_{\Pi^t} (U(\tfrac12)) ~.
\end{align}
As in \cite{GV} (see above the eq. (3.34) of the paper) we assume that
\begin{align}
 |\Lambda| + |\Xi| = |\Pi| ~, \label{assumegv}
\end{align}
which leads to
\begin{align}
 N_{\Lambda \Xi}^\Pi = \lim_{N \to \infty} N_{\Lambda \Xi}^{(2N)\Pi}
\end{align}
with $N_{\Lambda \Xi}^{\Pi}$ as the Clebsch-Gordan coefficient of gl$(\infty)_+$.
Using the fact that the Clebsch-Gordan coefficients are the same for the transposed representations,
we can show
\begin{align}
\begin{split} \sum_{\Lambda , \Xi} |  b^\lambda_{\Lambda;\Xi} (q)  |^2
  & =Z_\text{gauge}   \sum_{\Lambda , \Xi} |q^{\frac{\lambda}{2} (|\Lambda| - |\Xi|)}  \text{ch}_{\Lambda^t} (U(\tfrac12)) \, \text{ch}_{\Xi^t} (U(\tfrac12))|^2 \\
   & = Z_\text{gauge}   \sum_{\Lambda , \Xi} |  \text{ch}_{\Lambda^t} (U(h_+)) \, \text{ch}_{\Xi^t} (U(h_-))|^2 ~. \end{split}%\nonumber
\end{align}
This reproduces the gravity partition function \eqref{gpf}.
It was argued in \cite{Gaberdiel:2011zw,GV} that a large number of null states appear
in the 't Hooft limit and the decoupling of these null states is equivalent to the
condition \eqref{assumegv}. This assumption is quite important for the equivalence of the
partition functions, so we would like to examine it more carefully as a future problem.

\subsection{Characters from free fermions}

Here we would like to compute \eqref{a00} and \eqref{a0Xi} using free fermions.
It is well known that $\widehat{\text{so}}(2N)_1$ can be expressed by $2N$ real free fermions
$\psi^a$ with $a=1,2,\ldots , 2N$.
Thus the task here is to decompose the left hand side of \eqref{defa} by the characters
of the zero mode so$(2N)$ Lie algebra. The space of free fermions is spanned by
\begin{align}
 \prod_{j=1}^{n_\psi} \psi^{a_j}_{- r_j - \frac12} \Omega ~,
\end{align}
where $r_j = 0 , 1, 2 ,\ldots $ and $\Omega$ is the vacuum.
The branching function $a_{0;\Xi}$ counts the multiplicity when the representation
$\Xi$ of so$(2N)$ appears. The representation appears for the first time when
$n_\psi = |\Xi|$. Following the argument below (2.49) of \cite{CG}, the branching
function is found to be
\begin{align}
 \text{ch}_{\Xi^t} (U(\tfrac12))
 \label{factoru}
\end{align}
when summing the possible modes $r_j$ while keeping $n_\psi = |\Xi|$ fixed.
Notice here that the modes $r_j$ can be interpreted as the entries of the Young tableau of
the shape $\Xi$ since the Fermi statistic explains the rules for the entries (see figure \ref{fig:one}).

There is another contribution to the branching function  $a_{0;\Xi}$ with $n_\psi > |\Xi| $,
which comes from multiplying with so$(2N)$ invariants.
{}From the classical invariant theory, we can see that the so$(2N)$ invariants are
generated by \cite{We}
\begin{align}
 \prod_{r,s =0}^{\infty} (\sum_{a=1}^{2N} \psi_{-r- \frac12 }^a \psi_{-s - \frac12}^a)^{M_{rs}} ~
\end{align}
and additional invariants that have conformal dimension at least $N$. Note, that there are also only relations between invariants for
conformal dimension of that order due to the first and second fundamental theorems of invariant theory for the vector representation of the orthogonal group.
Now that we are dealing with Majorana fermions,
there are two differences from (2.51) of \cite{CG}. Firstly, we should
set $r \neq s$ due to the Fermi statistic. Secondly, we should set $r > s$  since two
fermions can be exchanged. In the large $N$ limit, we ignore the finite $N$ effects
and obtain
\begin{align}
 a_{0;0} (q) = \prod_{r > s=0}^{\infty} \sum_{M_{r,s}=0}^\infty
 q^{(r+s+1)M_{r,s}} = \prod_{r > s=0}^{\infty} \frac{1}{1 - q^{r+s+1}}
  =   \prod_{l=1}^\infty\prod_{n=2l}^{\infty} \frac{1}{1 - q^n }~,
\end{align}
which is \eqref{a00}. By multiplying with \eqref{factoru}, we obtain \eqref{a0Xi}.

\section{${\cal N}=1$ supersymmetric holography}
\label{susy}

In \cite{CHR} we have proposed the untruncated ${\cal N}=2$ supersymmetric version of the duality in \cite{GG}, and the
equivalence of the spectrum has been shown in \cite{CG}.
The aim of this paper is to conjecture an ${\cal N}=1$ version of the duality and show that
the one-loop partition functions agree.
The gravity theory considered is  the ${\cal N}=1$ truncation of the ${\cal N}=2$ higher spin
supergravity discussed in  subsection \ref{N=1g} (see also appendix \ref{N=1}).
We propose that the dual CFT is given  by the super coset model \eqref{susycoset}
\begin{align}
 \frac{\widehat{\text{so}}(2N+1)_k \oplus \widehat{\text{so}}(2N)_1}{\widehat{\text{so}}(2N)_{k+1}} ~.
 \label{scoset}
\end{align}
Moreover, we take a large $N,k$ limit while keeping the 't Hooft parameter \eqref{susypara}
finite, and identify the parameter with $\lambda$ appearing in the masses \eqref{massCHRB} and \eqref{massCHRF}.
Geometrically, the quotient by the right action of $\text{SO}(2N)$ is a sphere
\begin{align}
 S^{2N} = \frac{\text{SO}(2N+1)}{\text{SO}(2N)} ~.
\end{align}
Our coset, however, is a quotient by the adjoint action of $\text{SO}(2N)$.
Introducing $2N$ fermions, given by $\widehat{\text{so}}(2N)_1$ factor, the model has ${\cal N}=(1,1)$ supersymmetry.
This model should have the symmetry obtained by the Drinfeld-Sokolov reduction of
osp$(2N+1|2N)$ and this should be studied as a next step.
In the following we will show that the gravity partition function is reproduced by the
't Hooft limit of the ${\cal N}=(1,1)$ super coset \eqref{scoset}.

\subsection{The dual CFT}

We would like to compute the torus partition function of the coset model \eqref{scoset}
in the 't Hooft limit and compare it with the gravity partition function.
The central charge of the coset model is given by
\begin{align}
 c = \frac{k (2N+1) N}{k + 2 N - 1} + \frac{N (2N-1)}{2N-1} - \frac{(k+1) N (2N-1)}{k+2N -1}
    = \frac{3Nk}{k + 2N -1} ~.
\end{align}
The states are labeled as in the bosonic case. Namely, we use $\Lambda,\omega,\Xi$
as the highest weights of representations of $\widehat{\text{so}}(2N+1)_k$, $\widehat{\text{so}}(2N)_1$, $\widehat{\text{so}}(2N)_k$,
see appendix \ref{so} for some basics of the orthogonal Lie algebras.
The selection rule is now given by
\begin{align}
\Lambda + \omega - \Xi \in Q_{2N+1}
\label{sel}
\end{align}
where $Q_{2N+1}$ is the root lattice of so$(2N+1)$. For the identification of so$(2N+1)$ and so$(2N)$ weights, see again appendix \ref{so}.
Since the fermions in the gravity theory satisfy the anti-periodic boundary condition
along the space-like circle, we have to use the same boundary condition for the fermionic
states.  We set $\omega = \text{NS}$, where NS means the sum of the identity representation
$(\omega =0)$ and the vector representation $(\omega = 2)$.
With this $\omega$, the selection rule reduces to $\Lambda_N = \Xi_{N-1} + \Xi_{N}$ mod 2
with the notation in appendix \ref{so}.
The states are thus labeled by $(\Lambda;\Xi)$ with the field identification taken into account,%
\footnote{The field identification can be read off from the phases of character modular transformations
\cite{Gepner:1989jq}, and it may be written as
 $(\Lambda,\omega;\Xi) \sim (A \Lambda,\omega +2; \tilde A \Xi )$. Here $A$ is the $\mathbb{Z}_2$ outer
automorphism of $\widehat{\text{so}}(2N+1)_k$. The group of outer automorphisms of
$\widehat{\text{so}}(2N)$ is $\mathbb{Z}_4$ or $\mathbb{Z}_2 \times \mathbb{Z}_2$ depending on whether $N$
is odd or even. We set $\tilde A$ to be one of the four that exchanges $\Xi_0$ and $\Xi_1$ (and possibly
others), where the affine Dynkin labels  are represented as $[\Xi_0 ; \Xi_1 , \ldots , \Xi_{N}]$.
}
and
the conformal weights are
\begin{align}\label{eq:conformalweights}
 h_{(\Lambda ; \Xi)} = \frac{1}{k+2N-1} \left[ C_{2N+1} (\Lambda) - C_{2N}(\Xi) \right] + \frac{\omega}{4} + n ~.
\end{align}
Here $C_{M} (\Lambda)$ is the quadratic Casimir of so($M$) and
and the integer $n$ is the grade at which $\Xi$ appears in $(\Lambda , \omega)$.

We use the embedding $\text{SO}(2N) \hookrightarrow \text{SO}(2N+1)$ as
\begin{align}
 \imath (v) =
 \begin{pmatrix}
 1 & 0 \\
 0 & v
  \end{pmatrix}
  \in \text{SO}(2N+1) ~.
\end{align}
Then we embed $\widehat{\text{so}}(2N)_{k+1}$ diagonally into $\widehat{\text{so}}(2N)_k \otimes \widehat{\text{so}}(2N)_1$.
The states of the coset is obtained by the decomposition
\begin{align}
 \Lambda \otimes \text{NS} = \oplus_\Xi (\Lambda ; \Xi) \otimes \Xi ~.
\end{align}
The branching function of the coset model \eqref{scoset} can be defined as in the bosonic case
\begin{align}
 \text{ch}_\Lambda^{2N+1,k} (q, \imath(v)) \,  \text{ch}_\text{NS}^{2N,1} (q,v)
 = \sum_\Xi sb^{2N,k}_{\Lambda ; \Xi}  (q) \, \text{ch}_\Xi^{2N,k+1} (q,v) ~.
\end{align}
Here the character of the $2N$ free fermions in the NS-sector is
\begin{align}
\text{ch}_\text{NS}^{2N,1} (q,v) = \prod_{n=0}^\infty \prod_{i=1}^N (1 + v_i q^{n+\frac12})(1 + \bar {v_i} q^{n+\frac12}) ~,
\end{align}
where $v$ is an SO$(N,N)$ matrix with eigenvalues $(v_i,\bar v_i)$.
We consider the Hilbert space spanned as ${\cal H} = \sum (\Lambda; \Xi)
\otimes \overline{(\Lambda; \Xi)}$ and the partition function is
\begin{align}
{\cal  Z}^{2N,k} (q) = |q^{-\frac{c}{24}}|^2 \sum_{[\Lambda;\Xi]} | sb^{2N,k}_{\Lambda ; \Xi}  (q)|^2 ~.
\end{align}
Note that this is not the diagonal modular invariant in the usual sense, since we now take the
sum of the identity and the vector representations for $\widehat{\text{so}}(2N)_1$ in the numerator of \eqref{scoset}
in both the chiral and anti-chiral part.

In order to compare the CFT partition function with the supergravity partition function, we have to take
a large $N,k$ limit with the 't Hooft parameter \eqref{susypara}
\begin{align}
\lambda = \frac{2N}{k+2N-1}
\label{sthooft}
\end{align}
kept finite.
Now the states of the ${\cal N}=(1,1)$ super coset model \eqref{scoset} are labeled by
$(\Lambda ; \Xi)$, where $\Lambda, \Xi$ are the highest weights of so$(2N+1) $ and
so$(2N)$, respectively.
As mentioned in the bosonic case, the highest weight representations for so$(2N)$ can be labeled by
Young diagrams with finitely many boxes in the limit, and the quadratic Casimir becomes
$C_{2N}(\Xi) \sim N |\Xi|$ where $|\Xi|$ is the number of boxes
of Young diagram denoted by the same letter $\Xi$.
The case of so$(2N+1)$ is also studied in appendix \ref{so} and the same conclusions are obtained
in this case. Namely, the highest weights are described by Young diagrams in the limit, which
implies that the states of the super coset \eqref{scoset} are labeled by sets of two Young diagrams
$(\Lambda ; \Xi)$.
The quadratic Casimir is \eqref{casimirso2}
\begin{align}
 C_{2N+1}(\Lambda) = |\Lambda| N + \frac12 \left(\sum_{i=1}^{N-1} l_i^2 - \sum_j c_j^2 \right) ~,
\end{align}
where the number of boxes in the $i$-th row is $l_i$ and the number in the $j$-th column is $c_j$.
In the large $N$ limit, this behaves as $C_{2N+1}(\Lambda) \sim N|\Lambda|$.

We end this subsection by noting that the most fundamental states in the coset, which we believe to generate the remaining states under fusion in the 't Hooft limit, are $(v,0;0)$ together with its fermionic partner $(v,v;0)$, and $(0,0;v)$ also together with its fermionic partner $(0,v;v)$. Here $v$ denotes the vector representation with only the first Dynkin label non-zero and equal to one. Using the equation for the conformal weights \eqref{eq:conformalweights} we get
\begin{align}
 h_{(v,0;0)}&=\frac12\lambda\ ,& h_{(v,v;0)}&=\frac12(1+\lambda)\ ,\nonumber \\
 h_{(0,v;v)}&=\frac12(1-\lambda)\ ,& h_{(0,0;v)}&=\frac12(2-\lambda)\ ,
\end{align}
where in the last case we had to use $n=1$ in \eqref{eq:conformalweights} since we start from the trivial representation in the numerator of the coset. This fits perfectly with the calculated conformal dimensions from the bulk side \eqref{eq:bulkconformaldimension}. We will thus have two supermultiplets in the CFT generated from $(v,0;0)\otimes(v,0;0)$
and $(0,v;v)\otimes(0,v;v)$, respectively.

\subsection{Comparison of partition functions}

In order to compare the CFT partition function with its gravity dual, we have to take
the 't Hooft limit.
For large $k$, the leading terms of the characters are
\begin{align}
  &\text{ch}_\Lambda^{2N+1,k} (q,e^{\imath (H)}) \simeq
  \frac{q^{h_\Lambda^{2N+1,k}} \text{ch}_\Lambda^{2N+1} (e^{\imath (H)})}{\prod_{n=1}^\infty [ (1 - q^n)^N
    \prod_{\alpha \in \Delta_{2N+1}}( 1 - e^{\alpha (\imath(H))} q^n ) ] }  ~, \\
  &\text{ch}_\Lambda^{2N,k+1} (q,e^H) \simeq
  \frac{q^{h_\Lambda^{2N,k+1}} \text{ch}_\Lambda^{2N} (e^H)}{\prod_{n=1}^\infty [ (1 - q^n)^N
    \prod_{\alpha \in \Delta_{2N}}( 1 - e^{\alpha (H)} q^n ) ] }
\end{align}
as in \eqref{KW}, where $\Delta_{2N+1}$ and  $\Delta_{2N}$ denote the root systems of so$(2N+1)$ and so$(2N)$, respectively. As before, the Lie algebra characters appear in the limit.
The conformal dimensions are now
\begin{align}
 h_\Lambda^{2N+1,k} = \frac{C_{2N+1} (\Lambda)}{k+2N-1} ~, \qquad
 h_\Lambda^{2N,k+1} = \frac{C_{2N} (\Lambda)}{k+2N-1} ~.
\end{align}
Defining the leading term for large $k$ as
\begin{align}
 sb^{2N,k}_{\Lambda;\Xi} (q) \simeq
q^{h^{2N+1,k}_\Lambda - h^{2N,k+1}_\Xi}  sa^{2N}_{\Lambda;\Xi} (q) ~,
\end{align}
we have  $k$-independent relations
\begin{align}
  \text{ch}^{2N+1}_\Lambda (\imath (v)) \vartheta (q,v) = \sum_{\Xi} sa^{2N}_{\Lambda ; \Xi} (q) \,
  \text{ch}_\Xi^{2N} (v) ~.
  \label{srel}
\end{align}
The roots in $\Delta_{2N+1}$ include $\pm e_j$ $(j=1,\ldots N)$ in the orthogonal basis
in addition to those in  $\Delta_{2N}$, and from this fact we have
\begin{align}
 \vartheta (q , v) = \prod_{n=0}^\infty \prod_{i=1}^N
 \frac{(1+ v_i q^{n+\frac12}) (1+ \bar{v_i} q^{n+\frac12})}
        {(1- v_i q^{n+1}) (1- \bar{v_i} q^{n+1})} ~. \label{vartheta}
\end{align}
Setting $\Lambda = 0$ in \eqref{srel}, we obtain
\begin{align}
 \vartheta (q,v) = \sum_{\Xi} sa^{2N}_{0; \Xi} (q) \,
  \text{ch}_\Xi^{2N} (v) ~. \label{vthetaa}
\end{align}
If we use the decomposition
\begin{align}
 \text{ch}_\Lambda^{2N+1} (\imath (v) ) = \sum_\Phi R^{(2N)}_{\Lambda \Phi}
\text{ch}_\Phi^{2N} (v)
\end{align}
and \eqref{cgc}, then we find
\begin{align}
 sa^{2N}_{\Lambda ; \Xi} = \sum_{\Phi , \Psi} R^{(2N)}_{\Lambda \Phi} N_{\Phi \Psi}^{(2N) \Xi}
 sa^{2N}_{0;\Psi} (q) = \sum_{\Phi , \Psi} R^{(2N)}_{\Lambda \Phi} N_{\Phi \Xi}^{(2N) \Psi}
 sa^{2N}_{0;\Psi} (q) ~.
\end{align}

Now we take the 't Hooft limit. In the limit, the highest weights of so$(2N+1)$ and so$(2N)$ are
expressed by Young diagrams.
In the next subsection, we will find
\begin{align}
 & sa_{0;0} (q) = \lim_{N \to \infty} sa^{2N} _{0;0} (q)=\prod_{l=1}^\infty \prod_{n=2l}^\infty \left[ \frac{1}{1 - q^n} \right]^2
 \prod_{s=1}^\infty \prod_{n=s}^\infty ( 1 + q^{n+\frac12} )  ~, \label{sa00}\\
&  sa_{0;\Xi} (q) = \lim_{N \to \infty} sa^{2N}_{0;\Xi} (q)
= sa_{0;0} (q) \, \text{sch}_{\Xi^t} ({\cal U} (\tfrac12))  ~, \label{sa0Xi}
\end{align}
where the supercharacter is defined in \eqref{schL}.
The conformal dimensions are
\begin{align}
 h_\Lambda^{2N+1,k} \simeq \frac{\lambda}{2} |\Lambda| ~, \qquad
 h_\Xi^{2N,k+1} \simeq \frac{\lambda}{2} |\Xi| ~.
\end{align}
As in \cite{GV}, we assume the decoupling of null states in the 't Hooft limit,
which means that only the terms with
\begin{align}
 | \Phi| + |\Xi| = |\Psi|
 \label{assumes}
\end{align}
contribute to the fusion rules.
In the large $N$ limit, the coefficients stabilize as
\begin{align}
 \lim_{N \to \infty} N^{(2N) \Psi}_{\Phi \Xi} = N_{\Phi \Xi}^{\Psi} ~, \qquad
  \lim_{N \to \infty} R^{(2N) }_{\Lambda \Xi} = R_{\Lambda \Xi} ~,
\end{align}
where $N_{\Phi \Xi}^{\Psi} $ are the Clebsch-Gordan coefficients of gl$(\infty)_+$.
Just like the u$(M)$ case discussed in \cite{CG}, the restriction functions for the so$(M)$ case
become $R_{\Lambda \Phi} = N_{\Lambda |\Lambda/\Xi|}^{ \Xi}$ as shown in \cite{King:1975vf}.
Here $|\Lambda/\Xi|$ represents the Young diagram with a single row and with
$|\Lambda| - |\Xi| $ number of boxes.
Then eqs. (3.57) and (3.59) of \cite{CG}
\begin{align}
 \text{sch}_\Lambda \, \text{sch}_\Xi = \sum_\Xi N_{\Lambda \Xi}^\Pi \, \text{sch}_\Pi ~, \qquad
 \text{sch}_\Lambda ({\cal U } (0))
 = \sum_{\Xi} R_{\Lambda \Xi} \, \text{sch}_{\Xi^t} ({\cal U}(\tfrac12))  ~,
\end{align}
which were proven in appendix A of that paper, lead to
\begin{align}
sb^\lambda_{\Lambda ; \Xi} &= q^{\frac{\lambda}{2} (|\Lambda | - |\Xi| )}
 \sum_{\Phi , \Psi} R_{\Lambda \Phi }N_{\Phi \Xi}^\Psi \, sa_{0;0} (q)
\, \text{sch} _{\Psi^t} ({\cal U} (\tfrac12)) \nonumber \\
 &= sa_{0;0} (q) \, \text{sch} _\Lambda ({\cal U} (h_+)) \, \text{sch} _{\Xi^t} ({\cal U} (h_-)) ~.
\end{align}
Combining the anti-chiral part, we have
\begin{align}
 \sum_{\Lambda , \Xi} |sb^\lambda_{\Lambda ; \Xi} (q) |^2 = {\cal Z}_\text{gauge} \sum_{\Lambda , \Xi}
  |\text{sch} _\Lambda ({\cal U} (h_+)) \, \text{sch} _{\Xi^t} ({\cal U} (h_-)) | ^2 ~,
\end{align}
which reproduces the supergravity result \eqref{sgpf}.

\subsection{Characters from free bosons and fermions}

Let us now derive \eqref{sa00} and \eqref{sa0Xi} using free bosons and fermions.
The character \eqref{vartheta} is that of $2N$ real fermions $\psi^a$ $(a=1,2,\ldots , 2N)$
and $2N$ real bosons $J^c$ $(c=1,2\ldots , 2N)$.
The Fock space is spanned by
\begin{align}
 \prod_{j=1}^{n_\psi} \psi^{a_j}_{-r_j - \tfrac12} \prod_{l=1}^{n_J} J^{c_l}_{- t_l - 1} \Omega ~,
\end{align}
where $r_j,t_l$ run over non-negative integers.

{}From \eqref{vthetaa}, we can see that
the branching function $sa_{0;\Xi}$ counts the multiplicity when the representation
$\Xi$ of so$(2N)$ appears. The representation appears for the first time when
$|\Xi| = n_\psi + n_J$. Following the argument above (3.69) of \cite{CG}, the branching
function is found to be
\begin{align}
 \text{sch}_{\Xi^t} (U(\tfrac12))
 \label{sfactor}
\end{align}
when summing over the possible modes and $n_\psi,n_J$ while keeping the sum $n_\psi + n_J = |\Xi|$ fixed.
We also need to consider the so$(2N)$  invariant states
\begin{align}
 \prod_{r,s =0}^{\infty} (\sum_{a=1}^{2N} \psi_{-r- \frac12 }^a \psi_{-s - \frac12}^a)^{K_{rs}}
 \prod_{t,u =0}^{\infty} (\sum_{a=1}^{2N} J_{-t- 1 }^a J_{-u - 1}^a)^{L_{tu}}
 \prod_{r,u =0}^{\infty} (\sum_{a=1}^{2N} \psi_{-r- \frac12 }^a J_{-u - 1}^a)^{P_{ru}} ~.
\end{align}
Here we should set $K_{rs}$ non-zero only for $r > s$ and  $L_{tu}$ non-zero only for $t \geq u$.
We also set $P_{ru}=0,1$ for all $r,u$ since they are fermionic operators.
Invariant states can be constructed using Weyl's fundamental theorems of invariant theory for the orthogonal group \cite{We}.
Invariants that are not polynomials of the above states appear only for conformal dimensions larger than $N$, and also non-trivial relations
between invariant states appear for the first time at that order of conformal dimension.
In the large $N$ limit, we can ignore these finite $N$ effects
and obtain
\begin{align}
 sa_{0;0} (q) &= \prod_{r > s=0}^{\infty} \sum_{K=0}^\infty q^{(r+s+1)K}
  \prod_{t \geq u=0}^{\infty} \sum_{L=0}^\infty q^{(t+u+2)L}
   \prod_{r,u =0}^{\infty} \sum_{P=0}^1 q^{(r+u+\tfrac32)P}  \nonumber \\
 &= \prod_{r > s=0}^{\infty} \frac{1}{1 - q^{r+s+1}}
 \prod_{t \geq u=0}^{\infty} \frac{1}{1 - q^{t+u+2}}
 \prod_{r,u = 0}^{\infty} (1 + q^{r+u+\frac32} ) \nonumber \\
& =\prod_{l=1}^\infty \prod_{n=2l}^\infty \left[ \frac{1}{1 - q^n} \right]^2
 \prod_{s=1}^\infty \prod_{n=s}^\infty ( 1 + q^{n+\frac12} )
\end{align}
as in \eqref{sa00}. By multiplication with \eqref{sfactor}, we get \eqref{sa0Xi}.

\section{Symmetries of the dual conformal field theory}
\label{symmetry}

The symmetry algebra of the coset
\begin{align}
 \frac{\widehat{\text{so}}(2N+1)_k \oplus \widehat{\text{so}}(2N)_1}{\widehat{\text{so}}(2N)_{k+1}} ~,
\end{align}
is the commutant subalgebra $\text{Com}(\widehat{\text{so}}(2N)_{k+1},\widehat{\text{so}}(2N+1)_k \oplus \widehat{\text{so}}(2N)_1)$ of $\widehat{\text{so}}(2N+1)_k \oplus \widehat{\text{so}}(2N)_1$ that commutes with $\widehat{\text{so}}(2N)_{k+1}$.
Here, by $\widehat{\text{so}}(2N)_1$ we mean rank $2N$ free fermions $\mathcal F_{2N}$. Since they transform in the $2N$-dimensional vector (standard) representation
of $\text{so}(2N)$, they contain a homomorphic image of $\widehat{\text{so}}(2N)_1$ as subalgebra.
In addition, as we will see, the coset algebra is too large as it also contains additional fields with spin of order $N$. In order to get rid of these additional fields, one
needs an orbifold projection by improper orthogonal transformations.
At finite $N$, the candidate coset for the $\mathcal N=1$ super W-algebra is thus
\begin{align}
\text{Orb} \Bigl(\frac{\widehat{\text{so}}(2N+1)_k \oplus \mathcal F_{2N}}{\widehat{\text{so}}(2N)_{k+1}}\Bigr) ~,
\label{sorbifold}
\end{align}
but since the additional fields appear at conformal dimension at least $N$, these are invisible in the large $N$ limit, in other words, the spin content for the coset and its orbifold is the same for large $N$.

\subsection{The dimension $3/2,2,2,5/2$ fields of the coset algebra}

Before we discuss the complete coset algebra, we will explicitly compute the fields of lowest conformal dimension.
For this we need some preparation,
we write so$(2N+1)=\ $so$(2N)\oplus m$, where $m$ carries the standard representation of so$(2N)$.
We denote the currents of $\widehat{\text{so}}(2N+1)_k$ that belong to $m$ with the Roman indices $J^i,J^j,\ldots$
and those associated to so$(2N)$ with Greek indices $J^\alpha,J^\beta,\ldots$.
In addition, the fermions of $\widehat{\text{so}}(2N)_1$ also transform in the standard representation and we denote them by
$\psi^i,\psi^j,\dots$.
Then, the operator products expansions are
\begin{equation}
\begin{split}
J^\alpha(z)J^\beta(w) \ &\sim \ \frac{k\delta_{\alpha,\beta}}{(z-w)^2}+\frac{f^{\alpha\beta\gamma}J^\gamma(w)}{(z-w)} \, ,\\
J^\alpha(z)J^i(w) \ &\sim \ \frac{f^{\alpha ij}J^j(w)}{(z-w)} \, ,\\
J^i(z)J^j(w) \ &\sim \ \frac{k\delta_{i,j}}{(z-w)^2}+\frac{f^{ij\alpha}J^\alpha(w)}{(z-w)}\, .\\
\end{split}
\end{equation}
We also need the operator product with normal ordered products of currents. Denote by $(J^\alpha J^\beta)(z)$ the normal ordered product of two currents, then we e.g. have
\begin{equation}
\begin{split}
J^\alpha(z)(J^\beta J^\gamma)(w)\ \sim \ &\frac{kf^{\alpha\beta\gamma}}{(z-w)^3}+\frac{k\delta_{\alpha,\beta}J^\gamma(w)+k\delta_{\alpha,c}J^\beta(w)+f^{\alpha\beta\delta}f^{\delta\gamma \epsilon}J^\epsilon(w)}{(z-w)^2}+\\
&+\frac{f^{\alpha\beta\delta}(J^\delta J^\gamma)(w)+f^{\alpha\gamma\delta}(J^\beta J^\delta)(w)}{(z-w)}\,.
\end{split}
\end{equation}
With the help of this formula, we compute
\begin{equation}
\begin{split}
J^i(z)(J^\alpha J^\alpha)(w)\ &\sim\ \frac{(2N-1)J^i(w)}{(z-w)^2}+\frac{f^{i\alpha j}((J^jJ^\alpha)(w)+(J^\alpha J^j)(w))}{(z-w)} \, , \\
J^\alpha(z)(J^iJ^i)(w)\ &\sim\ \frac{4J^a(w)}{(z-w)^2} \, ,\\
J^i(z)(J^jJ^j)(w)\ &\sim\ \frac{(2k+2N-1)J^i(w)}{(z-w)^2}-\frac{f^{i\alpha j}((J^jJ^\alpha)(w)+(J^\alpha J^j)(w))}{(z-w)} \, .\\
\end{split}
\end{equation}
The fermionic fields are denoted by $\psi^i$ with operator product expansion
\begin{equation}
\psi^i(z)\psi^j(w) \ \sim \ \frac{\delta_{i,j}}{(z-w)} ~,
\end{equation}
and the corresponding currents are
\begin{equation}
j^\alpha\ = \ -\frac{1}{2}f^{\alpha ij}\psi^i\psi^j ~.
\end{equation}
Moreover  $\psi^i$ are primaries in the standard representation for these currents
\begin{equation}
j^\alpha(z)\psi^i(w) \ \sim \ \frac{f^{\alpha ij}\psi^j(w)}{(z-w)} ~.
\end{equation}
Let $\gamma=2k+4N-2$, then the Virasoro field of the coset algebra is
\begin{equation}
\begin{split}
T \ &= \ T_{\widehat{\text{so}}(2N+1)_k}+ T_{\text{fermion}}-T_{\widehat{\text{so}}(2N)_{k+1}} \\
&= \ \frac{1}{\gamma} \bigl((J^iJ^i)-2(J^\alpha j^\alpha)+(2k-3)T_{\text{fermion}}\bigr) ~.
\end{split}
\end{equation}
The coset symmetry algebra is the algebra that commutes with the $K^\alpha=J^\alpha+j^\alpha$, this is certainly true for
\begin{equation}
G\ = \ \sqrt{\frac{2}{\gamma}} (J^i\psi^i) ~.
\end{equation}
The fields $G$ and $T$ obey the operator product algebra of the $\mathcal N=1$ super Virasoro algebra, that is
\begin{equation}
\begin{split}
T(z)T(w)\ &\sim \ \frac{c/2}{(z-w)^4}+\frac{2T(w)}{(z-w)^2}+\frac{\partial T(w)}{(z-w)} \, ,\\
T(z)G(w)\ &\sim \ \frac{3G(w)/2}{(z-w)^2}+\frac{\partial G(w)}{(z-w)} \, ,\\
G(z)G(w)\ &\sim \ \frac{2c/3}{(z-w)^3}+\frac{2T(w)}{(z-w)} \, .\\
\end{split}
\end{equation}
Next, there is an additional dimension two field that commutes with both $j^\alpha$ and $J^\alpha$,
hence also with $K^\alpha$, and this is the Virasoro field $\tilde T$ of the coset
\begin{equation}\label{eq:boscoset}
\frac{\widehat{\text{so}}(2N+1)_k}{\widehat{\text{so}}(2N)_k}\, .
\end{equation}
Explicitly, in terms of currents, it reads
\begin{equation}
\begin{split}
\tilde T \ &= \ T_{\widehat{\text{so}}(2N+1)_k}-T_{\widehat{\text{so}}(2N)_{k}} \\
&= \ \frac{1}{\gamma} (J^iJ^i)-\frac{1}{\gamma(\gamma-2)}(J^\alpha J^\alpha) \, .
\end{split}
\end{equation}
This field is not a Virasoro primary, but the following linear combination
\begin{equation}
W_2\ = \ c\tilde T-\beta T\, ,\qquad\beta\ = \ \frac{4kN}{\gamma^2(\gamma-2)}((\gamma-2)(2k+2N-1)-4N+2)
\end{equation}
is. This statement is a straightforward computation using the above operator product expansions.
The dimension $5/2$ partner $G_{5/2}$ of $W_2$ can be computed as
\begin{equation}
\begin{split}
G(z)W_2(w)\ &\sim \ \frac{G_{5/2}(w)}{(z-w)} \, ,\\
G_{5/2}\ &= \ \frac{1}{\sqrt{2\gamma}} \Bigl(\beta(2(J^i\partial\psi^i)-(\psi^i\partial J^i))+a f^{ij\alpha}(((J^\alpha J^j)\psi^i)+((J^jJ^\alpha)\psi^i))\Bigr) \, ,\\
a\ &= \  \frac{12kN}{\gamma^2}((2k+2N-1)+2\frac{\gamma}{\gamma-2})\, .
\end{split}
\end{equation}
The operator product of $W_2$ with itself does not generate new fields, and it is
\begin{equation}
\begin{split}
 W_2(z)W_2(w)\ \sim \ &\frac{(c_b-\beta)^2c/2}{(z-w)^4}+\frac{2(c-2\beta)W_2(w)+2\beta(c-\beta)T(w)}{(z-w)^2}+\\
&+ \frac{(c-2\beta)\partial W_2(w)+\beta(c-\beta)\partial T(w)}{(z-w)}\ ,
\end{split}
\end{equation}
where $c_b$ is the central charge of the bosonic coset \eqref{eq:boscoset}.
The computation of further operator products becomes very complicated, but in a large $k$ limit they simplify just
as explained in section 6.5
of \cite{CHR2} for the case of the $\mathcal N=2$ coset.
We find
\begin{equation}
 \begin{split}
\lim_{k\rightarrow \infty} W_2(z)G_{5/2}(w)\ &\sim \ -\frac{6N^2G(w)}{(z-w)^3}-
\frac{2N^2\partial G(w)+NG_{5/2}(w)}{(z-w)^2} \\
&\hspace{+0.5cm} - \frac{\frac{N^2}{2} \partial^2 G(w)+\frac{2N}{5}\partial G_{5/2}(w)+ G_{7/2}(w)+\frac{3N}{2}A_{7/2}(w)-\frac{12N^2}{5}B_{7/2}(w)}{(z-w)} \ .
 \end{split}
\end{equation}
Here $G_{7/2}$ is a fermionic primary field of conformal dimension $7/2$, and $A_{7/2}$ and $B_{7/2}$ are dimension $7/2$ descendents that together with $G_{7/2},\partial G_{5/2}$
and $\partial^2 G$ form an orthogonal basis of dimension $7/2$ fields.
The precise form of these fields in the large $k$ limit is 
\begin{equation}
 \begin{split}
G_{7/2} &= \frac{3N}{2}\partial G_{5/2}+\frac{5N}{2}(W_2G)+\frac{3N^2}{5}\partial^2G-\frac{9N^2}{\sqrt{k}}(J^i\partial^2\psi^i)-4N(W_2G)+\frac{12N^2}{5}(TG)\, ,\\
A_{7/2}&= \frac{N}{2}(W_2G)+\frac{3N}{10}\partial G_{5/2}\, , \qquad B_{7/2}= (TG)-\frac{3}{8}\partial^2G.   
 \end{split}\, 
\end{equation}
We believe that the fields $G,T,W_2$ already generate the full symmetry algebra under iterated operator products.

\subsection{The field content of the coset algebra}

We now consider the field content of the coset algebra. Note, that the following analysis is in many respects similar to the one of last section, in particular, it relies on the classical invariant theory.

The generic field content of a coset algebra can under certain circumstances be computed using classical invariant theory
\cite{deBoer:1993gd}. The coset, we are interested in, is of this favourable type.
As mentioned before, the coset algebra is the commutant subalgebra
\begin{equation}
\text{Com}(\widehat{\text{so}}(2N)_{k+1},\widehat{\text{so}}(2N+1)_{k}\oplus  \mathcal F_{2N}) \, .
\end{equation}
The algebra $\widehat{\text{so}}(2N+1)_{k}\oplus  \mathcal F_{2N}$ is generated as a conformal field theory by the fields
$J^\alpha$ generating the $\widehat{\text{so}}(2N)_{k}$ subalgebra of $\widehat{\text{so}}(2N+1)_{k}$, the fields $J^i$ which are primaries in the vector representation
of $\widehat{\text{so}}(2N)_{k}$, and the fermions $\psi^i$.
An alternative set of generators is $K^\alpha,J^i$ and $\psi^i$, where the fields $K^\alpha$ generate the $\widehat{\text{so}}(2N)_{k+1}$ subalgebra of the commutant problem.
Note that $J^i$ as well as $\psi^i$ are primaries in the vector representation of $K^\alpha$.
In such a situation, it was argued that the fields of the commutant subalgebra are those that can be identified with the $\text{SO}(2N)$ invariant
products of the vector representation as follows.
Let
\begin{equation}
p(J,\partial J,...,\partial^m J,\partial \psi,...,\partial^m \psi)
\end{equation}
be a normally ordered polynomial in $J,\psi$ and their derivatives that is invariant under the natural action of $\text{SO}(2N)$. Then
\begin{equation}
p(J,\mathcal D J,...,\mathcal D^m J,\mathcal D \psi,...,\mathcal D^m \psi)
\end{equation}
is a generator of the coset algebra, and all generators are of such a form. Here, the covariant derivative is
\begin{equation}
\mathcal D I^i \ = \ \partial I^i +\frac{1}{k+1}f^{\alpha ij}(J^\alpha I^j) \, ,\qquad I^i\in\{J^i,\psi^i\}\,.
\end{equation}
We are thus left with determining all invariants in the vector representation. Weyl's first fundamental theorem for the orthogonal group \cite{We}
tells us that all such invariants are expressible in terms of the basic invariants which are traces of two vectors and determinants of matrices whose columns are
$2N$ vectors. Clearly, the determinants have spin at least $N$ and thus they are invisible in the large $N$ limit.
Also, note that the determinants are improper invariants, this means they change sign under transformations by orthogonal matrices with determinant minus one.
The traces are proper invariants and all proper invariants can be expressed in terms of traces \cite{We}.
We have three types of traces
\begin{equation}
A_{(n,m)}\ = \ \mathcal D^n J^i\mathcal D^m J^i,\qquad
B_{(n,m)}\ = \ \mathcal D^n \psi^i\mathcal D^m \psi^i,\qquad
C_{(n,m)}\ = \ \mathcal D^n J^i\mathcal D^m \psi^i.
\end{equation}
The spins are
\begin{equation}
\Delta(A_{(n,m)}) = n+m+2,\quad  \Delta(B_{(n,m)}) = n+m+1, \quad
\Delta(C_{(n,m)}) = n+m+\frac{3}{2}.
\end{equation}
Now, if there were no relations between products of the fields, then we can count that the algebra we found has a
generating set of fields whose bosonic fields
have spin $2,2,4,4,6,6,...$ while the fermionic generators have spin $3/2,5/2,7/2,...$.
Note, that these fields are  multiplets of the $\mathcal N=1$ super algebra as $(3/2,2),(2,5/2),(7/2,4),\ldots$.
The second fundamental theorem of invariant theory for the orthogonal group \cite{We} states that all relations between invariants either involve a determinant of a matrix whose
columns are vectors or they are determinants of $(2N-1)\times (2N-1)$ matrices whose entries are traces.
All these relations concern invariants whose spin is at least $N$ which implies that the spin content of the coset algebra agrees with the proposed higher spin supergravity in
the large $N$ limit.

We would like to remark, that the due to the determinants, the coset algebra at finite $N$ is larger than the $\mathcal N=1$ super W-algebra. In
\cite{Blumenhagen:1994wg} it has been proposed in an analogous situation to consider an orbifold in order to obtain a smaller coset algebra.
It is certainly possible that the invariant subalgebra, invariant under all improper orthogonal transformations, is the $\mathcal N=1$ super W-algebra.
As mentioned before, this is an issue which becomes invisible in the large $N$ limit and is thus of minor importance for the present purpose.

\section{Conclusion  and outlook}
\label{conclusion}

In this work, we propose that the $\mathcal N=1$ truncation of Prokushkin and Vasiliev's $\mathcal{N}=2$ higher spin supergravity on AdS$_3$ \cite{PV1} is dual to a limit of a family of conformal field theories given by the ${\cal N}=(1,1)$ super cosets
\begin{align}
 \frac{\widehat{\text{so}}(2N+1)_k \oplus \widehat{\text{so}}(2N)_1}{\widehat{\text{so}}(2N)_{k+1}} ~.
\end{align}
We need to take the large $N$ limit
with 't Hooft parameter
\begin{align}
 \lambda = \frac{2N}{2N + k -1}
\end{align}
kept finite. We have supported this conjecture by showing that
the supergravity  and the CFT partition functions match, i.e. the spectrum is the same on both sides. Further, we also studied the symmetry of the super coset model,
especially we provided explicit formulae for the fields of dimension $3/2,2,2$ and $5/2$.

It often happens that seemingly very different cosets possess the same symmetry algebra. We would like to remark that there are other cosets whose spin content of the symmetry algebra seems to coincide with the one of the coset of the present work.
For example consider
\begin{equation}
\frac{\widehat{\text{osp}}(1|2N)_k\oplus \mathcal B_N}{\widehat{\text{sp}}(2N)_{k-1/2}}\,.
\end{equation}
Here $\mathcal B_N$ denotes rank $N$ $\beta\gamma$-ghosts which contain as a subalgebra a homomorphic image of $\text{sp}(2N)_{-1/2}$.
The central charge of this coset is
\begin{equation}
c\ = \ -\frac{3Nk}{k+N+1/2}\,.
\end{equation}
This means that only for some negative levels $k$, we get a positive central charge.
The spin content of this coset can be studied as in the last section and with the help of Weyl's fundamental theorems of invariant theory for the symplectic group \cite{We}.
Indeed in the large $N$ limit, the spin content of this coset coincides with the spin content of the coset algebra studied in last section.
More cosets are constructed as follows. Let $S_{M|P}$ be the algebra generated by $2M$ free real fermions and $P$ $\beta\gamma$ ghosts, then $S_{M|P}$ contains a homomorphic image
of $\widehat{\text{osp}}(2M|2P)_1$ as subalgebra. The symmetry algebra of the cosets
\begin{equation}
\frac{\widehat{\text{osp}}(2M+1|2P)_k\oplus \mathcal S_{M|P}}{\widehat{\text{osp}}(2M|2P)_{k+1}}
\end{equation}
can be studied as before in the large $k$ limit and again seems to have the same spin content.
Conformal field theories of supercosets are usually not unitary and hence we expect
a tentative dual higher spin supergravity to be less interesting.

Further work is needed to obtain a better understanding of the duality.
In order to compare the partition functions, we assumed that some states in the CFT decouple from the
others in the large $N$ limit.  It is thus necessary to examine whether this assumption is true or not.
Also, it would be desirable to show that the asymptotic symmetry of the ${\cal N}=1$ supergravity
can be reproduced by the 't Hooft limit of the ${\cal N}=(1,1)$ super coset, in particular one should also study the
Hamiltonian reduction of the $\widehat{\text{osp}}(2N+1|2N)_k$ affine Lie algebra.
Furthermore, like in the cases of the other holographies on AdS$_3$, important checks of the duality would be to
calculate and compare correlators, and to consider the RG-flow.

\subsection*{Acknowledgements}

We are grateful to C. Candu, S.~Fredenhagen, M. Gaberdiel, K.~Ito, A.~Linshaw and V.~Schomerus for useful discussions.
The work of YH was supported in part by Grant-in-Aid for Young Scientists (B) from JSPS, and the work of PBR is funded by DFG grant no. ZI 513/2-1.

\appendix

\section{${\cal N}=1$ truncation of Prokushkin-Vasiliev theory}
\label{N=1}

The field equations of higher spin gravity theory by Prokushkin and Vasiliev can be expressed
in terms of generating functions $(W_\mu , B , S_\alpha)$ \cite{PV1}. Here $W_\mu$ is a space-time
one-form including the higher spin gauge fields, $B$ is a zero-form including matter fields, and $S_\alpha$ is an auxiliary field.
The generating functions depend on the parameters $(z_\alpha , y_\alpha ; \psi_{1,2}, k , \rho |x_\mu)$
where $x_\mu$ are the space-time coordinates.
The spinor index $\alpha$ takes values $1,2$.
The generating functions are expanded as
\begin{align}
\nonumber
 A (z , y ; \psi_{1,2} , k , \rho | x) = \sum_{B,C,D,E = 0}^1 \sum_{m,n = 0}^\infty
  A^{BCDE}_{\alpha_1 , \ldots , \alpha_m , \beta_1 , \ldots , \beta_n}
  k^B \rho ^C \psi_1^D \psi ^E_2 z^{\alpha_1} \ldots  z^{\alpha_m}
y^{\beta_1} \ldots y^{\beta_n} ~.
\end{align}
The Grassmann parity $\pi = 0 , 1$ is determined by the number of spinor indices as
\begin{align}
 & \pi (W_{ \alpha_1 , \ldots , \alpha_m , \beta_1 , \ldots , \beta_n}  ) =
 \tfrac12 ( 1 - (-1)^{|m+n|} ) ~,  \qquad
 \pi (B_{\alpha_1 , \ldots , \alpha_m , \beta_1 , \ldots , \beta_n}  ) =  \tfrac12 ( 1 - (-1)^{|m+n|} )  ~,  \nonumber \\
 & \pi (S_{ \alpha_1 , \ldots , \alpha_m , \beta_1 , \ldots , \beta_n}  ) =  \tfrac12 ( 1 - (-1)^{|m+n+1|} ) ~.
\end{align}
Moreover, we define a map $\sigma$ by
\begin{align}
 \sigma [A(z,y;\psi_{1,2},k,\rho)] = A^\text{rev} (-iz,iy;\psi_{1,2},k,\rho)
\end{align}
where the order of all generating elements is reversed in $A^\text{rev}$.
As shown in \cite{PV1} the following transformation is a symmetry of the field equations
\begin{align}
\eta (W_\mu) = - i^{\pi (W)} \sigma (W_\mu) ~, \qquad
\eta (B) = i^\pi (B) \sigma (B) ~, \qquad
\eta (S_\alpha) = i^{\pi (S)+1} \sigma (S_\alpha) ~.
\label{etat}
\end{align}
We can thus consistently truncate the fields to those invariant under this transformation, and this gives us
the $\mathcal{N}=1$ supersymmetric theory.

We consider vacuum solutions with $B = \nu$. In \cite{PV1} they obtained
three types of vacuum solution for $S_\alpha$, but we chose $S_{\alpha,0}^\text{sym}$ in
eq. (6.6) of that paper. The vacuum solution $W =W_0$ depends only on $(\tilde y_\alpha ; \psi_1,k)$. Here
$\tilde y_\alpha = \tilde y_\alpha^\text{sym}$ is defined in eq. (6.11) of \cite{PV1}, but all we need is that they obey
the following fundamental commutator
\begin{align}
 [ \tilde y_\alpha , \tilde y_\beta ] = 2 i \epsilon_{\alpha , \beta} (1 + \nu k ) ~,
 \qquad \{ \tilde y_\alpha , k \} = 0 ~.
\end{align}
Explicit forms of $S_{\alpha,0}^\text{sym}$ and $\tilde y_\alpha^\text{sym}$ will not
be used, but the following properties are important
\begin{align}
 \sigma[ S_{\alpha,0}^\text{sym}] = - i S_{\alpha,0}^\text{sym} ~, \qquad
 \sigma[ \tilde y_\alpha^\text{sym}] = i \tilde y_\alpha^\text{sym} ~.
\end{align}
In particular, $S_{\alpha,0}^\text{sym}$ is invariant under the action of $\eta$ defined in eq. \eqref{etat}.
Defining $A,\bar A$ as
\begin{align}
 W_0 = - \frac{1 + \psi_1}{2} A  - \frac{1 - \psi_1}{2} \bar A ~,
\end{align}
the field equations for $A, \bar A$ are given by those of the Chern-Simons theory
for the algebra generated by $(\tilde y_\alpha , k)$. This algebra was
called shs$[\lambda]$ algebra in \cite{CHR} where $\lambda = (1- \nu)/2$.
If we consider the sub-sector with even number of $\tilde y_\alpha$ and $k=1$,
then shs[$\lambda$] is reduced to its bosonic sub-algebra hs[$\lambda$].
The generators of shs[$\lambda$] may be given by \cite{CHR2}
\begin{align}
  V_m^{(s)+} = \left( \frac{-i}{4}\right)^{s-1} S^s_m ~, \qquad
  V_m^{(s)-} = \left( \frac{-i}{4}\right)^{s-1} S^s_m k ~, \qquad
  V^{(1)-} = k + \nu
\end{align}
with $s=2,3,\ldots$ for bosonic generators and $s=3/2 , 5/2 , \ldots$
for fermionic generators. Here $S^s_m$ is the symmetric product of $\tilde y_\alpha$s,
where $2s -2$ is the number of $\tilde y_\alpha$ and $2m = N_1 - N_2$ with
$N_{1,2}$ being the number of $\tilde y_{1,2}$.
For the bosonic generators, even $s$ generators are invariant under the action of \eqref{etat},
and for the fermionic generators,  $V_m^{(s)\pm}$ generators with
$s= 2n \mp 1/2 $ $(n=1,2,\ldots)$ survive.

We may define a different basis for the bosonic generators as
\begin{align}
 U_m^{(s)\pm} = \left( \frac{-i}{4}\right)^{s-1} S^s_m \frac{1 \pm k}{2}  ~.
\end{align}
Without the ${\cal N}=1$ truncation,  $U_m^{(s)+}$ generate hs[$\lambda$]
while $U_m^{(s)-}$ generate hs[$1-\lambda$].
It is known that the infinite dimensional Lie algebra hs[$\lambda$] can be truncated at
$\lambda = \pm n$ with integer $n $ and the reduced algebra becomes sl$(n)$ \cite{Fradkin:1990qk}.
In the same way, the even spin sub-algebra of hs[$\lambda$] is reduced at $\lambda = \pm n$
to sp$(n)$ for even $n$ and so$(n)$ for odd $n$ (see, e.g., \cite{GV}).
Thus, the ${\cal N}=1$ truncation of shs$[\lambda]$ can be reduced at
$\lambda = 2N+1$ to a superalgebra whose bosonic sub-algebra is given by
$\text{so}(2N+1) \oplus \text{sp}(2N)$.
Notice that $(V^{(2)+}_{m},V^{(3/2)+}_{r})$ with $m=0,\pm 1$ and $r=\pm 1/2$ are
the generators of the osp$(1|2)$ sub-algebra.
In terms of the superprincipal embedding of osp$(1|2)$, the generators of osp$(2N+1|2N)$
can be decomposed by the representation of osp$(1|2)$. The action of
$V^{(3/2)+}_{\pm 1/2}$ produces fermionic generators from bosonic ones in
the same representation of osp$(1|2)$.
This implies that the ${\cal N}=1$ truncation of shs$[\lambda]$ can be reduced
to the \mbox{$\text{osp}(2N+1|2N)$} superalgebra
since the bosonic sub-algebra of  osp$(2N+1|2N)$ is $\text{so}(2N+1) \oplus \text{sp}(2N)$.
In other words, the symmetry for
the massless gauge sector of the ${\cal N}=1$ truncated theory is given by an analytic continuation
of $\text{osp}(2N+1|2N)$ with $\lambda = 1 + 2N$. Or put differently, when $\lambda$ takes integer values, the $\mathbb{Z}_2$ automorphism $\eta$ defined in \eqref{etat} becomes the $\mathbb{Z}_2$ automorphism acting on supermatrices by a combination of (minus) supertransposition and
conjugation with a special matrix and this defines osp$(2N+1|2N)$ in terms of gl$(2N+1|2N)$, see \cite{Frappat:1992bs}.

The small perturbation by matter fields can be obtained by setting $B = \nu + {\cal C}$.
Studying the dynamical parts of ${\cal C}$, we can read off the matter content.
The ${\cal N}=1$ truncation of the matter fields is discussed around eq. (10.8) in \cite{PV1},
and it is given by an ${\cal N}=1$ hypermultiplet with two complex scalars
having masses
\begin{align}\label{}
 ( M^B_-)^2 = - 1 + \lambda  ^2 \ , \qquad( M^B_+ )^2 = - 1 + (\lambda - 1 )^2 ~,
\end{align}
 and two fermions with mass
\begin{align}\label{}
 (M^F_\pm)^2 = (\lambda - \tfrac{1}{2})^2 ~.
\end{align}

\section{Orthogonal Lie algebras}
\label{so}

Some basics of so($2N$) and so($2N+1$) Lie algebra are summarized.

\subsection{so($2N$) Lie algebra}

It will be convenient to introduce an orthogonal basis $e_i$ $(i=1,2,\ldots N)$ with
$e_i \cdot e_j = \delta_{ij} $.
In this basis, the roots of the so($2N$) Lie algebra are of the form $\pm e_i \pm e_j$ $(i \neq j)$
and the simple roots are
\begin{align}
 \alpha_i = e_i - e_{i+1} ~  (i = 1, \ldots , N-1) ~, \qquad \alpha_N = e_{N-1} + e_N  ~.
\end{align}
The fundamental weights are
\begin{align}
& \lambda_i = \sum_{l=1}^i e_l ~(i=1,\ldots, N-2) ~,  \nonumber \\
& \lambda_{N-1} = \tfrac12  ( e_1 + e_2 + \cdots + e_{N-1} - e_N ) ~, \\
& \lambda_{N} = \tfrac12  ( e_1 + e_2 + \cdots + e_{N-1} + e_N ) ~, \nonumber
\end{align}
and the Weyl vector is
\begin{align}
 \rho = \sum_{i=1}^N \lambda_i = \sum_{i=1}^N (N-i) e_i ~.
\end{align}

We consider a representation with the highest weight
\begin{align}
 \Lambda = \sum_{i=1}^N \Lambda_i \lambda_i ~,
\end{align}
where the coefficients are the Dynkin labels $\Lambda_i \geq 0$.
In the orthogonal basis, the highest weight can be expressed as
\begin{align}
 \Lambda = \sum_{i=1}^N l_i e_i ~.
\end{align}
with
\begin{align}
 l_i = \sum_{l=i}^{N-2} \Lambda_l + \tfrac12 (\Lambda_{N-1} + \Lambda_N) ~(i=1,\ldots N-2) ~,\\
 l_{N-1} = \tfrac12 (\Lambda_{N-1} + \Lambda_N ) ~, \qquad
 l_N = \tfrac12 (\Lambda_{N-1}- \Lambda_N) ~. \nonumber
\end{align}
In this basis, the quadratic Casimir is computed as
\begin{align}
 C_{2N}(\Lambda) = \tfrac12  \Lambda \cdot ( \Lambda + 2 \rho ) = \tfrac12 \sum_{i=1}^N l_i^2
  + \sum_{i=1}^N l_i (N - i) ~.
\end{align}
As shown in \cite{GV}, the quadratic Casimir for the representation with $\Lambda_{N-1} \neq 0$
and/or $\Lambda_{N} \neq 0$ is of order $N^2$ and these representations will be neglected.
In other words, we set $\Lambda_{N-1} = \Lambda_N =0$. Then the highest weight is
labeled by a Young tableau with $l_i$ boxes in the $i$-th row.
Notice that now $l_i \geq l_{i+1}$ and $l_{N-1} = l_N = 0$.
Denoting the number of
boxes in the $j$-th column by $c_j$,
 we have (see (A.9) of \cite{Gaberdiel:2011zw})
\begin{align}
 \sum_{i} i l_i = \frac12 \sum_j c_j ^2+ \frac{|\Lambda|}{2}
\end{align}
where the total number of boxes is denoted by $|\Lambda|$.
The quadratic Casimir is now
\begin{align}
 C_{2N}(\Lambda) = |\Lambda| \left(N-\frac12 \right) + \frac12 \left(\sum_{i=1}^{N-2} l_i^2 - \sum_j c_j^2 \right) ~,
 \label{casimirso}
\end{align}
and thus $C_{2N}(\Lambda) \sim N|\Lambda|$ in the large $N$ limit.

\subsection{so($2N+1$) Lie algebra}

We use the orthogonal basis $e_i$ $(i=1,\ldots, N)$ as in the so($2N$) case.
The roots of the so($2N+1$) Lie algebra are $ \pm e_j$ in addition to
$\pm e_i \pm e_j$ $(i \neq j)$  with $i,j = 1, \ldots , N$.
The simple roots are
\begin{align}
 \alpha_i = e_i - e_{i+1} ~  (i = 1, \ldots , N-1) ~, \qquad \alpha_N =  e_N  ~,
\end{align}
and the fundamental weights are
\begin{align}
 \lambda_i = \sum_{l=1}^i e_l ~(i=1,\ldots, N-1)  , \qquad
 \lambda_{N} = \tfrac12  \sum_{l=1}^N e_l  ~.
\end{align}
The Weyl vector is now
\begin{align}
 \rho = \sum_{i=1}^N (N + \tfrac12 -i) e_i ~.
\end{align}
The Dynkin labels $\Lambda_i \geq 0$ are the coefficients of the highest weight
\begin{align}
& \Lambda = \sum_{i=1}^N \Lambda_i \lambda_i = \sum_{i=1}^N l_i e_i ~, \\
 &l_i = \sum_{l=i}^{N-1} \Lambda_l + \tfrac12 \Lambda_N ~(i=1,\ldots N-1)~, \qquad
 l_N = \tfrac12 \Lambda_N ~.
\end{align}
The quadratic Casimir for the highest weight representation is
\begin{align}
 C_{2N+1}(\Lambda) =  \tfrac12 \sum_{i=1}^N l_i^2
  + \sum_{i=1}^N l_i (N + \tfrac12 - i) ~.
  \label{casimirso2}
\end{align}

It is easy to see that all the elements of the inverse of Cartan matrix
$C^{-1}_{ij} = \lambda_i \cdot \lambda_j$ are non-negative.
Thus we have
\begin{align}
 C_{2N+1} (\Lambda) \geq C_{2N+1} (\Lambda^{(s)}) =   N^2 \cdot \frac{a}{4} +
 \frac{a^2}{8} ~,
\end{align}
where $\Lambda^{(s)}_{i}=0$ for $i=1,\ldots , N-1$ and $\Lambda^{(s)}_N = a$.
This implies that $C_{2N+1}(\Lambda)$ is of order $N^2$ for representations with $\Lambda_N \neq 0$
as in the so($2N$) case, so we again set $\Lambda_N = 0$.
The highest weight representation is now labeled by a Young tableau with $l_i$ boxes
in the $i$-th row.
We denote the number of
boxes in the $j$-th column by $c_j$,
The quadratic Casimir is now
\begin{align}
 C_{2N+1}(\Lambda) = |\Lambda| N + \frac12 \left(\sum_{i=1}^{N-1} l_i^2 - \sum_j c_j^2 \right) ~,
\end{align}
which again leads to $C_{2N+1}(\Lambda) \sim N|\Lambda|$ for large $N$.

\end{document}